\documentclass[preprint1,eps]{aastex}
\usepackage{emulateapj5}
\input epsf

\slugcomment{To appear in {\it The Astrophysical Journal, Main Journal}, 558, Sept. 2001}

\shorttitle{Lobel \& Dupree}
\shortauthors{Kinematic Structure of Betelgeuse's Chromosphere}

\begin{document}

\title{Spatially resolved STIS spectroscopy of $\alpha$ Orionis: Evidence for non-radial chromospheric 
oscillation from detailed $\rm modeling^{1}$}

\author{A. Lobel and A. K. Dupree}
\affil{Harvard-Smithsonian Center for Astrophysics, 60 Garden Street, Cambridge MA 02138 \\
{\rm (alobel@cfa.harvard.edu, adupree@cfa.harvard.edu)}}

\footnotetext[1]{Based in part on observations with the NASA/ESA {Hubble Space Telescope} obtained at 
the Space Telescope Science Institute, which is operated by AURA, Inc., under NASA contract NAS 5-26555.}

\begin{abstract}
Four spatially resolved near-UV raster scans across the chromospheric disk 
of $\alpha$~Ori, obtained with the Space Telescope Imaging Spectrograph
on the {\it Hubble Space Telescope} reveal mean chromospheric infall from 1998 January to
1998 April, which reversed to upflow in deeper layers between 1998 September 
and 1999 March. In 1998 September we detect systematic reversals in the component 
maxima of four double-peaked emission lines of Si~{\sc i} (UV~1), Fe~{\sc ii} (UV~36), Fe~{\sc ii} (UV~61), and 
Al~{\sc ii}] (UV~1), when scanning across the UV disk. Detailed modeling of 
the Si~{\sc i} $\lambda$2516 resonance line with radiative 
transport calculations in spherical geometry constrain the mean 
radial velocity structure in the projected slit area (25 by 100~mas) for different 
aperture positions, observed off-limb to 157.5~mas. Hence we determine with semi-empirical models that   
these spatial reversals of emission line components correspond with average opposite flow velocities 
of $\sim$2~$\rm km\,s^{-1}$ across the chromospheric disk. 
We determine that the chromospheric velocity field can not 
be represented by a unique radial velocity structure across the stellar disk in 
order to match the observed peak ratios of this raster scan. 
These sub-sonic velocities indicate (local) non-radial  
movements of chromospheric fluid in confined regions during a chromospheric oscillation phase, which reverses 
from global contraction into expansion over this monitoring period of 15 months.
\end{abstract}


\section{Introduction}
Spatially resolved spectroscopic observations of gas dynamics in the atmospheres of 
stars other than the Sun are physically limited for various reasons, although they represent  
one of the most challenging measurements to improve our comprehension of stellar environments. 
Doppler information from ground-based optical high-resolution 
spectroscopy of spatially resolvable targets is restricted by the quality of seeing 
conditions through Earth's atmosphere, typically to the level of one arcsecond
in `good weather' conditions. Kinematic and morphologic 
studies of high-velocity winds from hot supergiants
indicate axis-symmetric nebulae \citep[i.e.][]{pas99},
although remnants from asymmetric mass ejection have also been observed \citep{mea99}.
Spectroscopic mapping studies of the low-speed winds from cool supergiants  
are scarce due to a natural deficiency of sufficiently nearby targets with resolvable optically bright  
circumstellar environments. \citet{mau90} detected profile changes in 
the spatially resolved K~{\sc i} (fluorescent) resonance emission line of $\alpha$~Ori (M2~Iab, HD39801), 
obtained off-limb at 5\arcsec and 7.5\arcsec\, from a scattering gas (and dust) shell  
which extends to at least 50\arcsec\, \citep{hon80} around this cool supergiant.    
\citet{maq90} also scanned the Na~{\sc i}~$D$ lines of $\alpha$~Ori and $\mu$~Cep (M2 Iae) 
at 6\arcsec, 10\arcsec, and 12\arcsec\, above the limb. These emission profiles reveal 
variable substructures for different slit positions and orientations. For $\mu$~Cep they are
interpreted to result from (incomplete) circumstellar shells with different expansion velocities.
Telluric Na~{\sc i} doublet emission at the short-wavelength side of these line profiles has 
been suggested to explain their 
double-peaked appearance in $\alpha$~Ori. We note that the detailed profiles of these optically 
thick lines bear however a striking resemblance to the self-absorbed line structure of many optically thick 
(fluorescent) emission lines of neutral and singly-ionized metals, observed with IUE and HST-GHRS
(Goddard High Resolution Spectrograph) in the near-UV spectrum of $\alpha$~Ori 
\citep[e.g.][]{bra95, car97}. 

The advent of space-borne spectroscopy has recently enabled high-resolution measurements at 
the subarcsecond level. In 1995 March, \citet{uit98} scanned the chromospheric disk 
of $\alpha$~Ori with GHRS and observed substantial changes in the prominent emission profiles of the Mg~{\sc ii}
UV resonance doublet. The aperture center scanned in perpendicular directions with successive 
steps of 27.5 mas, which is the apparent radius of the stellar disk in the optical. 
They found that the chromospheric emission from these lines reaches a diameter of 
at least 270 mas, about twice the size of the UV continuum (of ca. 120~mas), simultaneously obtained  
with the Faint Object Camera (FOC) by \citet{gil96}. These spectra 
were however obtained through an aperture with a projected dimension of 200 mas square,
which did not provide true spatial resolution on the chromospheric disk. 

A further reduction of the slit size down to 25 mas by 100 mas is currently offered with the 
Space Telescope Imaging Spectrograph (STIS). The brightness of $\alpha$~Ori in the near-UV is sufficient 
to provide the first truly spatially resolved spectra within realistic exposure times. 
Due to its proximity and supergiant size, Betelgeuse represents an exceptional target to
study the kinematic structure (i.e. the spatial and temporal properties) of an extended supergiant 
chromosphere at the limits of modern technological capabilities. 
These new measurements also represent a further increase of the spectral resolution in the raster scans, 
from medium-resolution with GHRS to $R$$=$114,000 with STIS. 
Detailed profile changes of chromospheric emission lines can be observed with high 
spatial and spectral resolution. We present a study 
of selected emission line profiles in the near-UV of this cool prototypical supergiant, 
from which we infer properties of the chromospheric velocity structure with detailed radiative 
transport calculations. This spatial and temporal modeling places direct constraints 
on the chromospheric kinematics between 1998 January and 1999 March. 
Section~2 presents a brief description of the observations. In Sect.~3
we compare and discuss the variable morphology of unsaturated 
self-absorbed emission profiles across the UV disk. In Sect.~4
we model these profile changes for the Si~{\sc i} resonance line, with simultaneous
downfalling and upflowing radial velocity structures observed in 1998 September.
We discuss chromospheric turbulence and oscillation inferred from this detailed modeling.   
The conclusions are presented in Sect.~5.   

\section{Observations}
A description of the STIS spectral scans used for the present study, 
with their wavelength and flux calibration, is provided in \citet{lob00}. 
For convenience we summarize the main properties of this data of HST proposal No.~7347.     
A spatial scan in the wavelength range between 2275~\AA\, and 3120~\AA\, was obtained 
on 1998 January~9, April~1, and September~24, and on
1999 March~28. The chromospheric disk was observed with an aperture size of 25 mas by 
100~mas, by stepping in the cross-dispersion direction with the medium-resolution grating.
This provided seven spatially resolved echelle spectra ($R$$\sim$30,000) in off-sets of 25 mas
across the disk, over a single scan axis almost parallel (within $10^{\circ}$) with the East-West direction.
The scan axis of 1998 January was tilted by $-$$38.9^{\circ}$ from E-W. Note that the stepping direction 
for the scan of 1998 September is reversed due to the $180^{\circ}$ roll in the seasonal 
orientation of the HST. The designation of Target Position (TP) from intensity peakup for this scan
is therefore spatially reversed for comparison with the other scans.
An overview of the scan axes\footnote{For the positions of the slit with  
respect to images, simultaneously obtained in the UV with the FOC, see 
http://cfa-www.harvard.edu/cfa/ep/pressrel/alobel0100.html.}
with respect to the North direction in the plane of the sky is given in Sect.~4.2.3.
The NUV-MAMA echelle line spread function and the absolute wavelength calibration 
for these spectra provide an accuracy within 1 pixel or 5~$\rm km\,s^{-1}$. The signal to noise 
ratios (S/N) of the individual spectra vary across the disk because different (single) exposure times 
were applied. These ratios typically range from 40 at intensity peakup to 25 near the disk edge. 
The data of 1998 April were complemented by an additional scan with the 
high-resolution grating ($R$=114,000) in the wavelength range of 2666~\AA\, to 2942~\AA. 
These observations required a bigger slit size of 63 mas by 200 mas, and the aperture 
center was placed in pointing offsets of 63~mas and 126~mas (left and right of intensity peakup). 
The wavelength calibration accuracy within one pixel corresponds to 1.3~$\rm km\,s^{-1}$.            
       
\section{Chromospheric emission line profiles}

\subsection{Line formation}

Numerous prominent emission lines from neutral and singly-ionized metals have been 
identified in the far- and near-UV spectrum. \citet{car94}
identified many blended Fe~{\sc ii} and S~{\sc i} lines between 1200~\AA\, 
and 2000~\AA. These Fe~{\sc ii} lines also dominate the spectrum from 2200~\AA\, to 2800~\AA\, 
\citep{bra95}, resulting from the complex term 
structure of Fe~{\sc ii}, which produces many blends between these lines, and 
with those of other elements. These profiles appear `double-peaked' because 
a central absorption core forms when the transitions become sufficiently 
optically thick in the chromospheric line formation region. For instance, Fe~{\sc ii} lines 
of several subordinate UV-multiplets (i.e. UV 60$-$64) with large oscillator strength, 
display strongly intensity saturated self-reversals with central depressions 
below the intensity level of the line wings and the background continuum. 
On the other hand, the lines with lowest log($gf$)-values in these multiplets show 
single-peaked emission profiles in which self-absorption is hardly discernible with high 
spectral resolution and S/N [i.e. $\lambda$2874.8 (UV~61), 
$\lambda$2716.7 (UV~62), $\lambda$2772.7 (UV~63)]. 

Central self-reversals form when the line source function is 
not monotonically decreasing with distance from the base of the chromosphere,
provided that the column density for the transition is sufficiently high
(due to line opacity resulting from the chromospheric extension and thermodynamic conditions). 
This scattering core of optically thick emission lines removes photons, 
produced in the deeper chromosphere, from the line of sight in the upper
layers where the source function decreases.
The frequency dependent contribution functions to the relative emission intensity 
in the line profile are therefore strongly depth dependent and  
determined by the detailed velocity structure in the line formation region. 

These complex profiles can be computed with radiative transfer through a 
well-constrained thermodynamic model of $\alpha$~Ori's extended chromosphere. 
A mean photospheric/chromospheric model of radial kinetic gas temperature, electron density, 
and microturbulence has been presented by \citet{lob00}, 
based on an elaborate fit procedure to the disk-integrated H$\alpha$ profile, the Ca~{\sc ii} K line, and 
the relative intensities and widths of the Mg~{\sc ii} UV resonance doublet.
This model reproduces the remarkable asymmetry observed for the latter
lines. The intensity of the blue $k$ emission component is partly reduced by a 
deep and narrow scattering core of the Mn~{\sc i} blend at 2794.8~\AA.
The frequency dependent contribution functions show how the line wings form 
below the central scattering core (see their Fig.~6). 
The relative contribution to the core flux increases in higher chromospheric layers  
with respect to the wings, which form deeper and closer to the 
chromospheric temperature rise. The emergent emission line profile is therefore 
dependent on the velocity structure in the formation region of this scattering core. 
Photons moving toward the observer in a collapsing chromosphere encounter a larger opacity 
at the long-wavelength side of the line than the short-wavelength, and hence 
the blue wing appears stronger and the red wing fainter \citep{hum68}.
A simple (parameterized) modeling of such asymmetric emission profiles of $\alpha$~Ori 
was carried out by \citet{boe79}, based on Fe~{\sc ii} lines with stronger 
blue components, observed around 3200~\AA\, in 1970$-$75.

The component maxima of these double-peaked metal emission lines are observed 
to reverse over time. \citet{bra95} observed reversals in Fe~{\sc ii}
$\lambda$2391.5 (UV~35), suggesting chromospheric inflow for 1988 March and outflow in 1989 January.
Inflow was again observed in 1991 January (IUE), whereas both components had equal intensity 
in 1992 September (GHRS). The latter spectrum was analyzed by \citet{car97} 
who inferred a correlation between the radial velocity of Fe~{\sc ii} central reversals  
and computed line opacities. They detected an acceleration of the mean chromospheric 
velocity structure from rest to $\sim$$-$7~$\rm km\,s^{-1}$. The present STIS monitoring 
also reveals how these cores displace over time (by 4 to 8 $\rm km\,s^{-1}$), which
correlates with the strengthening of the red emission component when this core
displaced bluewards between 1998 April and 1999 March \citep[see Fig. 7 of][]{lob00}.
These observations point to recurrent upflows in the stellar chromosphere 
which may be linked with the 400$-$420 day variability period found by \citet{dup87} 
from $U$-band photometry and Mg~{\sc ii} fluxes, and by \citet{smi89} from 
photospheric radial velocity variations. A short review of chromospheric variability 
based on earlier spectral data is also given in \citet{que86}.

\subsection{Spatial and temporal line component behavior}
Figure 1 shows the STIS raster scans obtained with the medium-resolution grating 
for our monitoring program. The profile changes observed in Fe~{\sc ii} $\lambda$2868.874 
(UV~61) ({\it dash-dotted lines}), Si~{\sc i} $\lambda$2516.12 (UV~1) ({\it solid lines}), and
Fe~{\sc ii} $\lambda$2402.6 (UV~36) ({\it dotted lines}) are shown in 
off-sets of 25 mas from intensity peakup (TP 0.0) on the chromospheric 
disk. Note that the Fe~{\sc ii} $\lambda$2868.9 line intensity has been scaled up by a factor 
of 3 for comparison with the Si~{\sc i} profiles. These lines are investigated because 
their self-absorption cores are not intensity saturated. 
Intensity saturation is observed in $\alpha$~Ori for the stronger 
lines of these UV multiplets. It distorts the wing profiles of the reversals,
thereby masking small changes in the Doppler position of this core. This is 
for instance observed in the wings of the wide and zero-flux self-absorption cores 
of Mg~{\sc ii} $h$ \& $k$, which remain static between 1998 January and 1999 March.  

In general, the scans of 1998 in Fig.~1 display stronger blue than red emission components, 
indicating downflow across the entire chromosphere. A closer inspection 
reveals that the component intensities are however not identical when scanning
left and right of peakup position. For instance, in 1998 April the blue component 
exceeded the red much more at negative scan positions than for positive 
off-sets. The component ratio for both lines is noticeably larger at TP $-0.025$ than at $+0.025$.
This difference results from the minimum in the Fe~{\sc ii} self-reversal at TP $+0.025$, 
which occurs $\sim$5~$\rm km\,s^{-1}$ blue-shifted with respect to TP $-0.025$. 
But the FWHM for the central core at TP $+0.025$ is also larger, which lowers the resulting 
intensity of its adjacent emission maxima. Although a similar asymmetry of component ratios 
is observed for the Si~{\sc i} line, clear differences in core position and width remain
undetectable, which results from the deeper central core of this 
resonance line, which dips about 10~$\rm km\,s^{-1}$ bluewards of the Fe~{\sc ii} centroid.

When scanning further off the disk in 1998 April we observe how the self-reversal
of the Fe~{\sc ii} $\lambda$2868 line becomes very weak at TP $-0.075$. A single-peaked
emission profile appears, whereas the profile at TP $+0.075$ remains double-peaked.  
To test the influence of instrumental broadening we compare the profiles with 
the high-resolution scan. This raster scan was obtained immediately after the 
medium-resolution scan, but with a 63~mas by 200~mas aperture. The high-dispersion profiles 
of Fe~{\sc ii} $\lambda$2868 are shown in Fig.~1 by the dashed drawn (noisily) lines. They correspond 
with the aperture placed off-limb at 0~mas (intensity peakup), $\pm63$~mas, and $\pm126$~mas. 
For comparison purposes we overplot these profiles at TPs $\pm0.050$ and $\pm0.075$, 
but recall that they sample a larger disk area and a wider 
spatial range. The absolute intensities are also scaled down for
an effective comparison with the medium-resolution profiles.
The component ratios in the high-resolution scan closely match (to within a few percent)
the ratios of the medium-resolution scan. 
Notice also the component asymmetry at $\pm63$~mas, and 
the same shape difference of a single emission line at $-126$~mas (shown at TP $-0.075$), but which is 
double-peaked at $+126$~mas (TP $+0.075$). This comparison reveals that the component ratios in the 
medium-resolution scans can be used to infer reliably the spatial structure of the 
chromospheric velocity field, because these emission lines are 
intrinsically sufficiently broad. The separation of both components exceeds 
40~$\rm km\,s^{-1}$ (or at least 4 times $\Delta$$\lambda$$\simeq$0.09~\AA), so that the difference
between $R$$\sim$30,000 and $R$=114,000 has no appreciable effect on the observed line shapes.
 
The medium-resolution scan of 1998 September in Fig.~1 reveals that the asymmetry 
of component ratios at opposite sides of the chromospheric disk increases further. 
The blue component of the Si~{\sc i} line ({\it solid lines}) at TP $+0.025$ diminishes 
to below the intensity of its red emission component. This reversal is observed for both 
the Si~{\sc i} and Fe~{\sc ii} lines, and extends farther out to TP $+0.075$. 
It suggests a reversal of the mean flow direction in the formation region of the self-absorption 
core, causing slight Doppler shifts of this core which strongly alter the emission maxima. 
In the next section we model these spatial reversals with small differences
in the mean chromospheric velocity structure of the exposed disk area. This expanding trend, first detected  
at positive scan positions (West of peakup), is further observed in the raster scan of 1999 March. 
For this epoch the intensity maximum of the red component of the iron lines strongly exceeds the blue one 
for all positive scan positions, including TP 0.0. In the meantime, the intensity excess  
in the blue components has strongly decreased at the negative scan positions.
This variability points to ongoing changes in the local chromospheric kinematics, 
extending further Eastwards across the chromospheric disk. 
 
Similar temporal and spatial evolutions are observed in Fig.~2 for other emission lines of 
Fe~{\sc ii} $\lambda$2917.416 (UV~61) ({\it dotted lines}) and Al~{\sc ii}] $\lambda$2669.166 (UV~1) ({\it solid lines}). 
The latter semiforbidden line also shows the spatial reversal of the component maxima
when scanning from TP 0.0 to $+0.025$ in 1998 September. 
In 1999 March both lines display a strengthening of the red peak over the blue maximum at TP 0.0.
However, component variations of lines that belong to the same multiplet are not necessarily identical. 
For example, Fe~{\sc ii} $\lambda$2880.756 ({\it long dash-dotted lines}) is a transition with large oscillator strength 
of the same $61^{\rm st}$ UV multiplet. This line permanently displays a stronger 
blue emission maximum (but is also blended with Cr~{\sc ii} $\lambda$2880.863  which could distort its shape). 
Differences in profile changes are also observed among other lines with high and low log($gf$)-values 
that belong to the same multiplet. In general, the leading lines of a multiplet have deep and broad self-reversals, 
and display stronger blue emission maxima during our monitoring period. On the other hand, the unsaturated 
lines display a systematic evolution, which comprises spatial and temporal intensity reversals of their component
maxima. Such lines are useful to track the detailed chromospheric kinematics. 
The component separation of core-saturated (i.e. strongly opacity broadened) lines of many 
Fe~{\sc ii} multiplets typically exceeds 60~$\rm km\,s^{-1}$, whereas the maxima of lines with unsaturated cores 
can be separated by less than 40~$\rm km\,s^{-1}$ (see Fig.~2). 
We note however that unblended lines with central absorption intensities above the 
level of the background continuum 
are rather scarce in the very crowded near-UV spectrum of $\alpha$~Ori. Two other examples are 
Fe~{\sc ii} $\lambda$2759.336 (UV~32) and Fe~{\sc ii} $\lambda$2449.739 (UV~34). These lines display morphologic 
changes very similar to the profiles shown in Fig.~1, including the sudden reduction of their 
component ratios from TP 0.0 to $+0.025$ in 1998 September. Similar behavior is observed in Fig.~2 for the 
somewhat stronger Fe~{\sc ii} $\lambda$2984.831 line of UV~78 ({\it short dash-dotted lines}). 
Note that this line displays a stronger red component in 1999 March, which is also observed 
at TP~0.0 for Fe~{\sc ii} $\lambda$2917 ({\it dotted lines}) and for the Al~{\sc ii} line ({\it solid lines}).
The weakly intensity saturated Fe~{\sc ii} $\lambda$2402 line in Fig.~1 ({\it dotted lines}),
with higher optical depth, also displays the spatial component reversal of 1998 September, which is observed for 
the two unsaturated lines in this Figure, and for the Al~{\sc ii} line of Fig.~2.

In Fig.~3 we show the reversals between TPs 0.0 and $+$0.025 for these four lines. 
The Al~{\sc ii} line ({\it solid lines in upper panels}) is drawn for two overlapping echelle orders. 
The close agreement confirms that instrumental noise remains small. Current STIS 
performance monitoring indicates that the E230M and E230H gratings are only 
affected by scattered light to 4\% at 2500~\AA\, \citep{sts99}.
Errorbars on the line intensities are also plotted for the Fe~{\sc ii} $\lambda$2869 line 
({\it dash-dotted lines in upper panels}), and the weaker Si~{\sc i} line ({\it solid lines in lower panels}). 
The errorbars are provided by the STIS calibration pipeline and are mainly determined by the accuracy 
of the absolute sensitivity calibration of the E230M grating and of the aperture throughput, 
and by the bias and background subtractions in case of low intensities. We observe these 
reversals however near disk center for which the spectra have good S/N$\sim$40, with global 
count rates $\sim$2000~$\rm s^{-1}$. The spatial intensity changes in the emission components 
exceed these errorbars, which indicates that the systematic differences result from changes 
in the line formation conditions across the chromospheric disk.   
  
In summary, we observe systematic profile changes in weak emission lines of various species, 
that are formed in Betelgeuse's chromosphere. The line shapes are characterized by asymmetries 
which vary with the chromospheric kinematics. Spatially resolved observations of selected (unblended and
unsaturated) lines reveal that complex chromospheric velocity structures are present
during these variability phases.    

\section{Chromospheric modeling}

\subsection{LTE synthesis}

The upper panel of Fig.~4 shows a portion of Betelgeuse's near-UV spectrum ({\it thick solid lines}), 
observed between 2735~\AA\, and 2750~\AA. This medium-resolution spectrum was 
obtained at intensity peakup (TP 0.0) in 1998 September. The region between 2736~\AA\, and 2745~\AA\,
is obtained from two overlapping echelle orders, for which the emergent fluxes closely match. 
This region is typical for the star's UV spectrum which displays a forest of emission peaks, 
crossed by broad zero-flux absorption troughs that are seen against an apparent mean level of the UV background. 
These zero-flux cores have been identified in \citet{bra95}, 
and result from prominent self-reversals of many blended Fe~{\sc ii} lines, 
mostly of UV-multiplets 62 and 63. 

The synthetic spectrum for this region is shown by the thin drawn solid line. 
For this computation we apply the mean thermodynamic model atmosphere of \citet{lob00}. 
This model has been determined from NLTE fits to the H$\alpha$ profile, observed in 1993 February. 
The normalized core depth of H$\alpha$ was equal to within 5\% of the depth observed in 1998 October. The spectrum in Fig.~4 is 
however computed for LTE conditions in plane-parallel geometry. This is because a detailed 
NLTE synthesis would require much more elaborate computations that solve the radiative transfer and rate equations 
for the very complex Fe~{\sc ii} atom. This is outside the scope of our present study which 
merely tries to identify the strongest spectral features from these approximate fits. We find that an LTE 
synthesis already meets this goal very closely. This spectral region is well-matched by only 
considering Fe~{\sc i}~- and Fe~{\sc ii}-lines in the synthesis, shown by the dotted line 
({\it upper panel}). The positions of the strongest Fe~{\sc ii} lines are marked
in the stellar rest frame. The lower energy levels of these lines do not exceed $\sim$3~eV, with 
log($gf$)-values typically above $-$1. The iron spectrum fits the majority of the 
observed strong spectral features. The synthetic spectrum computed with all  
atomic species is shown by the thin solid line. Differences with the iron spectrum remain limited, 
although some weaker absorption cores in the observed spectrum are also identified as 
Cr~{\sc ii} lines at the indicated wavelengths. The line list and the atomic line data
are available from Kurucz website (1998)\footnote{URL at http:/kurucz/linelists.html}. 
Note that the current atomic data produces unobserved strong iron emission lines, for example 
Fe~{\sc i} $\lambda$2737.633 and Fe~{\sc ii} $\lambda$2744.897. 
The log($gf$)-values of the transitions at $\lambda$2737 and at $\lambda$2745 
are computed to be too strong (respectively $-$1.801 and $-$1.727).
Note however that the small log($gf$)-value of $-$0.939, given in \citet{fuh88} 
for the Fe~{\sc ii} $\lambda$2741.394 line, produces an emission line intensity which matches the 
observations. 

On the other hand, our spectral synthesis correctly reproduces the strong self-absorption cores 
observed in the leading lines of subordinate iron multiplets (i.e. Fe~{\sc ii} $\lambda$2739.548 with 
log($gf$)=0.233), which suggests that these $f$-values are appropriate. The broad overlapping line wings 
occasionally match the observed near-UV background fluxes (i.e. near 2748~\AA). 
Note that \citet{car94} confirmed with GHRS spectra that 
this background, first detected with IUE, is a true continuum which is not due to scattering 
inside the spectrograph. They also argued that it appears to originate in the chromosphere 
at temperatures ranging from 3000$-$5000~K, in agreement with our semi-empirical chromospheric model.

It is of note that the near-UV fluxes of Betelgeuse also vary slightly 
over different regions of the inner chromospheric disk. \citet{gil96} found excess emission 
in the radial intensity distribution observed in the near-UV in 1995 March. This 
resulted from a brighter feature seen in the SW disk quadrant. This `spot' was observed within 25~mas of 
the disk center. Similar, but not identical, patterns in the UV-continuum brightness are also 
observed within $\sim$30~mas of intensity maximum, in dithered FOC images that are simultaneously monitored 
with the STIS spectra \citep{dup99}. We conjecture that these UV-continuum patterns 
are related to the asymmetric brightness distributions observed by \citet{you00} 
with optical interferometry. They suggest that the dominant TiO opacity in the optical 
is very sensitive to small differences in local thermodynamic conditions across the photosphere,
because no brightness asymmetries have been measured in the near-IR where TiO opacity is almost absent
(R. Kurucz, 2000, private communication). Probably, a similar temperature/density dependence for the chromospheric 
continuum level can alter the emergent near-UV fluxes that are observed across the chromospheric disk.  

The LTE spectral synthesis is performed through the stellar chromosphere to check for possible blends 
in the profile of the lines we discussed in the previous Section. 
The lower panels of Fig.~4 show the computed 
LTE spectrum near the Si~{\sc i} $\lambda$2516 line ({\it panel left}) and the Al~{\sc ii}] $\lambda$2669 
line ({\it panel right}). 
The observed profiles of 1998 September are boldly drawn. They are shown for TP 0.0 ({\it solid lines})
and TP $+$0.025 ({\it dashed lines}), and display the remarkable spatial reversal of their component maxima. 
Two Al~{\sc ii} profiles are drawn for each pointing position, obtained from two 
overlapping echelle orders. The computed profiles are shown by the thin
dash-dotted lines. The Si~{\sc i} resonance line becomes strongly 
self-absorbed because of its high log($gf$)-value (0.241) and very small lower energy level.
The Al~{\sc ii} is a zero electron-volt semiforbidden transition with a weaker self-reversal due 
to its smaller optical thickness and low log($gf$)-value of $-$4.69. 
After broadening the computed Si~{\sc i} profile with the instrumental resolution, 
and with a total value for $v$sin$i$ and macroturbulence  
of 9$\pm$1~$\rm km\,s^{-1}$, we find that the shape of the self-absorption core ({\it thin drawn solid line})  
closely matches the observed core profiles. The width of this scattering 
core is strongly dependent on the microturbulence velocity profile in the 
chromosphere, which ranges up to 19~$\rm km\,s^{-1}$ in our model. 

The component maxima of the computed profiles assume equal intensities because 
a hydrostatic model chromosphere is applied. This reveals that both lines are not
blended with other strong chromospheric emission lines. However, the spectral synthesis yields weak 
emission lines of Fe~{\sc ii} $\lambda$2516.407 and of Fe~{\sc i} $\lambda$2516.571. 
Both iron lines appear unobserved when comparing the computed (and broadened) profile 
with the long-wavelength wing of the observed profiles. Their log($gf$)-values 
are predicted to be too high, similar to the iron lines in the upper panel of this Figure. 
This is also found for the predicted Cr~{\sc ii} line at $\lambda$2668.707 with high log($gf$)=$-$0.524.
The strong emission components are not observed, but its deep self-reversal 
can be identified with a broad absorption core which is observed at the 
short-wavelength side of the Al~{\sc ii} line.

\citet{car97} measured highly supersonic average chromospheric turbulent velocities in the range 
of 31$-$35~$\rm km\,s^{-1}$ from the Doppler FWHM of intercombination lines of C~{\sc ii} and Si~{\sc ii}, 
and from fluorescent Co~{\sc ii} and Fe~{\sc ii} lines. These values are however based 
on the assumption that these emission lines can be considered as optically thin,
and that the broadening is controlled by turbulence rather than opacity effects.  
Our spectral synthesis and detailed radiative transfer modeling reveals however that 
macroturbulence is not the main broadening mechanism and that chromospheric emission 
lines in $\alpha$~Ori are considerably opacity broadened. Their high macrobroadening 
values are therefore strongly overestimated since the line widths are determined 
by various other broadening mechanisms. It is to be remembered that the FWHM of a spectral
line is determined by the thermal motions of the gas {\it and} the (projected) {\it micro}turbulence 
velocity when radiative transfer effects are important. Next to these effects, the 
lines shape can broaden further by {\it macro}broadening. The latter conserves the total
line flux, and can result from large-scale mass movements, stellar rotation ($v$sin$i$), 
and the instrumental dispersion. In the lower panel of Fig. 4 ({\it right}) we compute
that about half of the observed FWHM of the Al~{\sc ii} $\lambda$2669 intercombination line 
({\it bold lines}) results by radiative transport in the chromospheric formation region ({\it thin
dash-dotted line}), whereas the remainder of the line width can be ascribed to macrobroadening.       
Since the rotational broadening for Betelgeuse and the instrumental broadening are small
we compute a value of 9~$\rm km\,s^{-1}$ for the large-scale (macroturbulent) velocity 
for convolution with the predicted spectrum to obtain an overall best match with 
the observed near-UV spectrum ({\em upper panel}). 

Although the C~{\sc ii}] lines do not display self-absorbed central cores as the Al~{\sc ii}] line, 
it has been pointed out that these lines should {\it not} be considered as optically thin transitions.        
\citet{jud98} presented a discussion about effects of radiative transfer on the 
C~{\sc ii} intercombination lines. They assumed that these lines are `effectively' 
optically thin, and applied the `mean escape probability approximation' 
to determine the electron density in the mean formation region. They derive 
$N_{\rm e}$$\simeq$$10^{8.3}$ $\rm cm^{-3}$, which is about a factor of ten 
larger than in our semi-empirical model \citep{lob00}, which is based 
on detailed radiative transfer modeling of H$\alpha$, Mg~{\sc ii} $h$ \& $k$, and TiO bandheads. 
However, \citet{jud98} also conclude that the optical depth inferred from their line ratio estimates 
are high enough in the chromosphere to influence these ratios, and thereby 
``render all conclusions based upon the assumption of optically thin line formation 
invalid for these lines''. It has been known for some time that this assumption yields very 
different values for the chromospheric electron density. For example, 
\citet{ste81} measure $\sim$2$\times$$10^{8}$, $\sim$5$\times$$10^{7}$, and 
$\sim$1$\times$$10^{7}$ $\rm cm^{-3}$, for the three line ratios, assuming a temperature 
of $10^{4}$~K in the line formation region. Comparative 
studies by \citet{byr88} demonstrated that the relative intensities of the 
C~{\sc ii}] lines are variable between exposures with IUE. 
They noted that electron densities derived from different line ratios 
were incompatible, and reported evidence of variation in the line profile 
from one exposure to another. This variability points to line 
broadening mechanisms determined by important radiative transport effects in these 
transitions. 
   
\subsection{Detailed NLTE modeling}

The profile of the Si~{\sc i} $\lambda$2516 line also shows strong temporal changes besides 
the spatial variations discussed above. 
For instance, in 1992 the GHRS disk-integrated self-absorption core revealed a blue-shift 
by $\sim$10~$\rm km\,s^{-1}$, 
compared with the 1998-99 STIS scans. Its red component maximum strongly exceeded the blue one, 
while the former became much weaker in 1998-99. \citet{lob00} determined a mean chromospheric 
velocity structure from this line profile in 1992, which accelerates from rest to about $-$4~$\rm km\,s^{-1}$.    
The Si~{\sc i} profiles of 1998 January and April in Fig.~1 indicate inflow in the core formation region
since the blue component exceeds the red. But the spatial reversal 
observed at TP $+$0.025 in 1998 September (see Fig.~3) indicates the local alteration of this
downflow into upflow. We infer the corresponding kinematic structure in the chromosphere 
from a refined NLTE modeling of these variable component ratios, based on the thermodynamic 
model of \citet{lob00}. 

\subsubsection{Spatially resolved modeling}

We utilize the SMULTI code \citep{har94}. It computes emergent line
profiles in moving media and accounts for their formation in spherical geometry and non-LTE conditions. 
The equation of state considers the ionization stages of hydrogen and helium (for solar abundance 
values), and the electron pressure. The statistical equilibrium and rate equations 
are solved for two adjacent ionization stages of the species (with a user defined abundance)
for which we model the spectral line shape with detailed radiative transport calculations.
The code outputs the line flux, the ionization fraction for the considered element, together 
with its departure coefficients from LTE for the ionization and excitation fractions 
per model layer and atomic energy level. The non-LTE equilibria are very sensitive 
to the kinetic temperature and the local electron density ($N_{\rm e}$) adopted in the model atmosphere. 
The model is obtained by semi-empirically varying both structures with height, until 
the `best match' is obtained with the observed line flux and shape. The computation of 
emission lines formed in the chromosphere with spherical geometry is critically dependent 
on $N_{\rm e}$, because the chromosphere is predominantly cooled by mechanisms which depend 
linearly on $N_{\rm e}$ per gram \citep[electron collisional excitation of neutral and singly ionized 
species and $\rm H^{-}$ recombination; see][]{har94}.
The modeling method is based on the mass conservation technique \citep[Eq. (2) of][]{har94}, 
which yields slightly different temperature, gas-density, and -pressure structures for the different phases, with 
variable mass-loss rates. Thermodynamic models which also list various quantities like column mass, 
optical depth, mass density, number densities of H~{\sc i} and H~{\sc ii}, total gas pressure, 
electron pressure, turbulent pressure, pressure scale height, can be obtained in electronic form from the authors.

The term structure of the silicon atom is sufficiently simple (as opposed to iron), 
to converge the population densities throughout an extended model of the chromosphere, 
within acceptable runtimes. A Si model atom with 12 energy levels and the continuum is applied. 
We solve the radiative transfer and statistical equilibrium equations 
for 6 line transitions and 9 fixed continua. The line at $\lambda$2516
has a common upper level ($\chi_{\rm up}$=4.953~eV, $g$=5) with the lines at $\lambda$2506 and at 
$\lambda$2970 ($3s^{2}\,3p\,4s$ ${}^{3}P^{\rm o}$$\rightarrow$$3s^{2}\,3p^{2}$ ${}^{3}P$).
The former is a prominent emission line in the spectra, but the latter has not been 
identified, possibly due to blends with Fe~{\sc i} lines \citep{bra95}. 
Three other transitions connect the level at $\chi_{\rm up}$=4.929~eV ($g$=3) with the same lower levels of 
the above listed lines.
The intrinsically weakest line at $\lambda$2987 is observed as a strong emission line without a self-reversal.
The two other transitions at $\lambda$2519 and at $\lambda$2528 with oscillator strengths $\sim$20 times larger,
are strongly self-absorbed, and display weak emission components against the near-UV background.                  
\citet{car88} showed that the line at $\lambda$2516 (e.g. also observed in the M3 giant $\gamma$~Cru)
is a fluorescent line whose upper level is populated by a fluorescent Fe~{\sc ii} line at $\lambda$2506.8, 
which pumps the Si~{\sc i} line at $\lambda$2506.9. This could explain why the latter line appears so strongly
in emission, whereas the line at $\lambda$2519, with otherwise comparable oscillator strength and 
transition probability ($A_{ul}$), does not. The present modeling procedure does not readily 
account for changes in the branching ratios due to pumping, or by photon conversions due to this 
Fe~{\sc ii} blend. Such changes by extra photo-excitation on the emergent intensity ratios of these lines, 
sharing the same upper level, are however not relevant to fit {\em relative} intensities of the component 
maxima in the Si~{\sc i} $\lambda$2516 line. Fluorescence in this transition enhances both components equally, 
which can be accounted for by scaling up the predicted profile.

The upper left panel of Fig.~5 shows the best fit ({\it thin solid line}) to the 
Si~{\sc i} line observed at TP 0.0 in 1998 September ({\it bold solid line}).
The computed profiles in this Figure are spatially resolved by summing the intensity 
contributions from light rays that propagate parallel to the line of sight, and cross the projected 
surface area within the field of view, for each aperture position on the chromospheric disk.  
This integration is performed along the width (25~mas) and height (100~mas) of the slit, 
placed in off-sets of 25~mas from the center of our model atmosphere. 
These fluxes are multiplied by the geometric weighting factors to account for the spherical 
curvature of the atmosphere (or the path length for every ray within the sector of the chromospheric 
model included in the STIS aperture), to yield the emergent spatial line intensities. 
For these computations the photospheric core radius 
of 27.5~mas is set to $R_{\star}$=700~$R_{\odot}$, where $\tau_{5000}$=1. There are 21 model layers 
below this point (where $T$$\simeq$$T_{\rm eff}$), with the deepest layer  
located at 8.6$\times$$10^{11}$~cm ($\sim$12~$R_{\odot}$ or 0.017~$R_{\star}$) below the formation region of the 
photospheric continuum. The radiative transport is solved for 5 rays intersecting this core, 
whereas light rays that graze the photospheric core are computed throughout the outer 50 layers 
above the core radius. In these chromospheric layers the kinetic temperature $T$ 
increases from a minimum of 2769~K to a maximum of 5443~K, then decreases outwards to 
$\sim$3000~K at a height of 9.3~$R_{\star}$. For the {\it Hipparcos} \citep{esa97} distance of 131~pc to $\alpha$~Ori,
this height corresponds to $\sim$250~mas, which is beyond the outermost radius of  
$\pm$87.5~mas (at TPs $\pm$0.075) of our raster scan with the 25~mas aperture. Therefore, light rays that {\em start} 
from layers above 87.5~mas (outside the field of view) do not contribute to the intensity 
integration. These layers of course do contribute farther out to passing rays that start deeper, 
inside the projected slit area. Note that the pointing jitter for these observations in the HST 
fine-lock mode remains typically below 3.5~mas, which is sufficiently small with respect to the 
slit size to enable an effective comparison of the emergent line intensities with scan position.

The computed profile of Si~{\sc i} $\lambda$2516 at TP 0.0 in Fig.~5 matches the observed line shape with a
downfalling mean velocity structure, shown in the upper right-hand panel ({\it solid line}). 
We compute that a radial acceleration of this inflow, from rest to 1.4~$\rm km\,s^{-1}$ 
at 1.85~$R_{\star}$ above the photospheric radius, produces the observed intensity 
ratio of the line components within their errorbars at central intensity. An increase of the downflow velocity
by 1.6~$\rm km\,s^{-1}$ ({\it dotted line} for a terminal inflow velocity of 3~$\rm km\,s^{-1}$) 
increases this ratio by $\sim$25~\%, which indicates that the 
detailed Si~{\sc i} line shape serves as an accurate velocity indicator in the chromosphere. 
Note that there is slight movement of the central core depending on the 
terminal velocity, but the line asymmetry is more pronounced and more easily measured.
A velocity structure for which the inflow accelerates slowly ({\it long dash-dotted line}) to 1.4~$\rm km\,s^{-1}$
produces a blue emission component that is $\sim$20\% too weak compared to a velocity 
structure which steeply accelerates in the deeper chromospheric layers.      
The latter mean velocity structure provides the best match with the profile observed at TP~0.0. 
The contribution from the photospheric continuum becomes 
largest with respect to the line flux because the slit width mainly samples light rays that start 
inside the photospheric core. The rays that contribute along the slit height also include grazing 
rays out to 50~mas. On the other hand, the off-star scan positions chiefly integrate rays that graze 
the inner core. This adds little or no photospheric background continuum, which is shown in the lower 
panels of Fig.~5 by the thin drawn solid lines. Note that we compute this background with the Uppsala opacity package
in spherical geometry as well. The photospheric background flux is dominated by $bf$-opacity of H 
and $\rm H^{-}$, whereas in the chromosphere the continua of Al and Mg are important 
(i.e. the strong bound-free opacity edge of Mg~{\sc i} at 2513~\AA), 
besides the electron opacity. For a gas of solar abundance we compute that hydrogen is partially 
ionized up to 35~\% near the temperature maximum. 

Our detailed modeling procedure requires a redshift
for the synthetic profiles of $+$18.3~$\rm km\,s^{-1}$ to provide a best global
fit to the line shape observed at TP 0.0. The profiles are computed 
in the stellar rest frame, and have to be corrected for the stellar radial velocity
in the heliocentric frame. The computed profiles in Fig. 5 are plotted with a redshift
of $+$18.3~$\rm km\,s^{-1}$, for all the scan positions. It is of note 
however that this value does not equal the center-of-mass velocity of $\alpha$~Ori.
Between 1998 January and September we detect redshifts of $\sim$$+$8~$\rm km\,s^{-1}$ 
in the scattering cores of Fe~{\sc ii} $\lambda$2392 and Si~{\sc i} $\lambda$2516, 
which are typical for most emission lines in the near-UV spectrum \citep[][see Sect. 6 of]{lob00}. 
This results from mean chromospheric mass movements. For example, in 1992 September 
the central core of Si~{\sc i} $\lambda$2516 strongly blueshifted during a phase 
of global chromospheric expansion, and which enhanced its long wavelength 
emission component. Chromospheric emission lines exhibit temporal variations of 
radial velocity that are comparable to the amplitude of the photospheric radial velocity curve.        
An overview of optical radial velocity studies for $\alpha$~Ori is given in \citet{smi89}.  

These raster scans further reveal that the intensity of the background emission near 2500~\AA\, 
reduces almost linearly from TP~$+$0.025 to TP~$+$0.075, about halving at each subsequent scan position. 
This dimming is smaller in bright emission lines (or in their components) observed for this 
wavelength region. It results in brighter line emission 
maxima  (i.e. they show limb-brightening) with respect to the background level when scanning off the disk.
For the Si~{\sc i} line of Fig.~5 this ratio changes from $\sim$4 at TP~0.0 to 
$\sim$7 at TPs~$\pm$0.075. This distribution of emergent  
line intensities is well matched with our mean thermodynamic model atmosphere.
The model also reproduces the gradual weakening observed in the 
depth of self-absorption cores with respect to the off-star background levels. 
This spatial behavior is for instance 
clearly noticeable in the unsaturated core of the Fe~{\sc ii} line at 2868~\AA\, of Fig.~1. 
It results from the decrease in column density of the scattering material in the upper chromosphere 
with larger viewing angles. This effect is reproduced in Fig.~5 with our spatially resolved 
radiative transfer calculations. 

\subsubsection{Projection effect}

Another important effect in modeling Betelgeuse's chromospheric kinematics is the 
geometric projection of its (complex) wind structure when scanning towards the limb. 
For the ideal case of a radially expanding (contracting) wind, which sustains a single velocity 
with distance, the Doppler position of the self-reversal is displaced from shorter (longer) wavelengths
towards the rest-frequency of the line near the limb. Near the disk edge the central core forms in flow 
directions that are nearly perpendicular to the line of sight, whereas at disk center 
its formation chiefly samples the wind flow parallel with this line. Through this projection 
effect, emergent line profiles become more symmetric when scanning closer to the limb. 
The profile at peakup position (at or near the disk center) probes the radial component 
of the outflow (inflow) velocity because this self-absorption core mainly forms in front of the disk,
suppressing the resulting blue (red) emission component according the magnitude and direction 
of the mean flow velocity. 

In reality this ideal scenario is severely complicated in several ways. 
The size of the aperture still covers about a quarter of the UV-disk area, so that towards  
opposite ends of the slit, light rays are integrated that emerge from radial 
outflow (or inflow) with reversed directions, 
perpendicular to the line of sight on either side of the chromospheric front hemisphere. 
Although the intensity contribution from these rays is smaller 
than for rays collected near the middle of the aperture, they reduce the asymmetry of the integrated line profile. 
This effect diminishes when scanning off the disk where the field of view samples a more uniform (tangential) flow 
direction, but where the line intensity (and S/N) diminishes as well. Further complications arise from probable
velocity variations along the radial wind profile in the star's extended chromosphere. 
The detailed wind structure may accelerate and/or decelerate with distance above the photosphere. 
The higher inflow velocity at larger distance in Fig.~5 causes the strong component asymmetry 
computed for TP~0.0 ({\it thin solid line}), which is observed in 1998 April ({\it bold dash-dotted line}) and 
1998 September ({\it bold solid line}). For TPs~$\pm0.050$ and $\pm0.075$ the asymmetry in the computed profiles 
reduces noticeably due to the geometric projection. This effect is observed at the negative scan 
positions, in agreement with our spatially resolved modeling. But the profiles observed 
at positive scan positions clearly do not match this model with downflow.
These off-star component maxima in 1998 September reverse with respect to TP~0.0.
The emission maxima at the positive scan positions are best fit ({\it thin dash-dotted lines}) 
with an outflowing (mean) velocity of 0.5~$\rm km\,s^{-1}$, shown by the dash-dotted 
line in the upper right-hand panel. Note that it closely fits the relative and 
absolute component intensities at TPs $+$0.025 and $+$0.075.
However, the absolute intensity of the profile predicted for TP $+0.050$ is somewhat too strong.
This can result from the mean thermodynamic model we apply for all scan positions. 
Small spatial differences across the chromosphere in the local kinetic temperature or electron density 
may account for small differences in the absolute intensity distribution of the line.
This is not unexpected, since images of Betelgeuse display chromospheric surface features. 
On the other hand, the relative intensity distribution in the line components chiefly depends 
on the detailed chromospheric velocity structure, which is reproduced with this spatially 
resolved modeling. 

Further simulations reveal that the systematic and prominent component reversal from 
TP $0.0$ to TP $+0.025$ cannot be modeled with complex non-monotonic radial velocity structures 
that collapse and expand with depth. The reversals occur too close to the inner 
(photospheric) disk radius, because these profiles mainly sample chromospheric mass movements
parallel to our line of sight. This excludes the possibility that reversals can result from 
a combination of the projection effect and a {\it unique} but non-monotonic radial velocity structure.
In principle, such combination can produce component reversals for the outer 
scan positions, which sample a more uni-directional flow in the field-of-view.
However, for the inner scan positions appreciable contributions occur from light rays crossing 
opposite flow directions on either side of the front hemisphere, when a unique (but complex) radial 
velocity structure is assumed. Such contributions render the integrated profiles more symmetric,
contrary to what is observed at TPs 0.0 and $+$0.025. Our modeling shows that the reversals, observed at these
inner scan positions, can only be caused by an inversion in the mean radial flow direction between
these scan positions. This implies that the chromosphere assumes an inherently non-radial velocity 
structure in September 1998. 

\subsubsection{Chromospheric kinematics}

Figure~6 shows a three-dimensional schematic representation of the chromospheric velocity structure, inferred
from the spatially resolved observations and the detailed line profile modeling. The dashed lines 
mark the aperture positions with respect to the photospheric hemisphere, 
drawn by the dotted circles which contour surface areas of $30^{\circ}$ latitude and longitude. 
The size and direction of the arrows indicate the radial speed and flow direction 
above a surface area for the chromospheric fluid in the front hemisphere.
This radial velocity structure is represented at four depths in the chromosphere. 
It clearly illustrates the projection effect for grazing light rays that cross 
the chromosphere at different locations within the aperture, and for 
different scan positions of the latter on the chromospheric disk.     
The chromospheric velocity pattern of 1998 January shows a decelerating mean downflow 
since the blue component maxima of unsaturated lines are stronger across the disk.   
This observed line asymmetry slightly enhances for negative scan positions in 1998 April, 
which is represented by a somewhat stronger downflow on the Eastern (left-hand) hemisphere. 
The reversals discussed above in the component maxima of unsaturated lines, observed from TP 0.0 to $+0.025$ 
in 1998 September, are represented by outflow in the deeper chromospheric layers on the 
Western (right-hand) hemisphere (lower left graph). This local upflow occurs inside a larger 
collapsing chromospheric envelope. The mean downflow in this outer envelope is observed  
through the stronger blue components of many lines with saturated scattering cores  
(like Fe~{\sc ii} $\lambda$2881 and Fe~{\sc ii} $\lambda$2985 in Fig.~2) 
that sample a bigger and more extended chromospheric volume in the line 
of sight, with higher optical depths than the cores of the unsaturated lines. 
This global chromospheric inflow strongly decelerates or begins to reverse in 
1999 March when much more symmetric line profiles are observed. The 
reduction of the line asymmetry extends further towards the negative scan 
positions, and is represented by outflow from deeper layers, extending higher up 
in the chromosphere towards the Eastern hemisphere (lower right-hand graph). 
This systematic evolution indicates a stratified flow that is not strictly radial  
in Betelgeuse's extended chromosphere, and which reverses from a global contraction 
into expansion during this 15 months monitoring period.

\subsubsection{Chromospheric turbulence}  

The difference we determine for 1998 September of mean inflow of 1.4~$\rm km\,s^{-1}$ 
at TP 0.0 and mean outflow of 0.5~$\rm km\,s^{-1}$ at TP $+0.025$ is small, since it does 
not exceed the isothermal sound speed of $\sim$7~$\rm km\,s^{-1}$ in the chromosphere. 
However, the Doppler position of the Si~{\sc i} self-absorption core  
samples the depth-integrated fluid velocity. Local flow velocities 
in the chromosphere can be much higher on a length scale smaller than the mean formation region 
for this scattering core, but which become averaged out by this depth integral. This is illustrated by the small distance of 
only 0.015~$R_{\star}$ a `mean' upflow with 0.5~$\rm km\,s^{-1}$ would travel between 
1998 September and 1999 March. It is therefore relevant to compare these average 
fluid velocities with the large value for the macrobroadening of 9$\pm$1~$\rm km\,s^{-1}$, 
required to derive the correct line width from the predicted (NLTE) profile, after 
correcting for the instrumental spectral resolution. 
This value is also well-constrained because higher values would 
produce self-reversals that are too broad and too shallow for the detailed fits of Fig.~5. 
\citet{boe79} also determined that the observed profile shapes rule out 
large-scale non-thermal motion of more than 10~$\rm km\,s^{-1}$. 
Since the $v$sin$i$-value for this evolved supergiant is expected not to exceed $\sim$5~$\rm km\,s^{-1}$, 
this large macrobroadening-value can in part result from actual large-scale turbulent  
movements in the chromosphere. Strong large-scale up- and downflow in Betelgeuse's 
chromosphere cannot be ruled out because \citet{lob00} determined a value of 
12$\pm$0.5~$\rm km\,s^{-1}$ for the photospheric macrobroadening from single and unblended 
absorption lines observed with very high spectral resolution around 10800~\AA. 
These metal lines also provided an accurate value of 2$\pm$1~$\rm km\,s^{-1}$
for the projected microturbulence velocity in the photosphere. The large difference
between the photospheric microturbulence of $\sim$2~$\rm km\,s^{-1}$ and a macroturbulence velocity
of 16$\pm$9~$\rm km\,s^{-1}$, has also been observed for the 
M1.5 Iab supergiant $\alpha$~Sco by \citet{dek88}. 

The variable brightness patterns observed by \citet{dup99} 
in the near-UV of Betelgeuse can result from large-scale convective movements 
in the deeper photosphere which penetrate the lower chromosphere. 
Convection plumes that occasionally overshoot the chromospheric temperature 
minimum will cause small differences in the local kinetic temperature and produce the 
observed flux variations. Most importantly, our best fits to the line width 
observed near the disk center and near the limb in Fig.~5 require a constant value for 
the velocity of this `chromospheric macroturbulence', projected in the line of sight. 
For every scan position we convolve with the same Gaussian macrobroadening profile.
It indicates that these large-scale turbulent motions remain isotropic or angle independent  
across the chromosphere. This constant value for macrobroadening is also measured in 
the scan of 1998 January (see Fig. 1), when the scan axis was tilted by $-$$39^{\circ}$ 
with respect to the East-West direction. A clear change of the total line width caused 
by the different slit orientation is not apparent, which indicates that the observed 
macrobroadening is not determined by high projected rotation velocities 
($v$sin$i$-values) in this line. The spatial invariance for this broadening, 
observed in these optically thick emission lines, appears to be linked with an isotropic and uniform
large-scale velocity field in the star's extended chromosphere.
The limited (subsonic) changes we observe in the average fluid velocity across the chromosphere,  
and this invariable macro-turbulence field could indicate that these small velocities are  
interdependent or originate from the same type of velocity field. 
In other words, subsonic non-radial oscillations of the entire 
chromosphere may produce uniform (macroturbulent) eddies with similar 
or higher velocities on smaller length scales (e.g. like shaking a `snowglobe'). The latter are clearly 
not resolved with the 25 mas by 100 mas aperture since the observed line broadening remains  
constant across the disk. This observation strongly suggests that 
macro-turbulence in Betelgeuse's {\em chromosphere} cannot be interpreted as due to the spatial 
inhomogeneity of a large-scale granular velocity field that consists of a few convection 
cells.  

Energy cascading from large-scale eddies onto smaller length-scales with higher velocities, 
under gravitational confinement, is also consistent with the highly supersonic 
microturbulence velocities in our chromospheric model. These velocities are required to match the shape and 
width of the self-absorption core, or the separation of the emission line components (see Figs. 4 \& 5).
The core forms on average higher in the chromosphere than the emission line wings, 
where this small-scale velocity field also increases in our model. 
It is of note that \citet{dej97} have shown that the supersonic 
photospheric microturbulence observed for the hotter and more luminous supergiant
$\rho$~Cas (F2$-$G $\rm Ia^{+}$) can be mimicked with hot and thin sheet-like layers, 
produced by the wakes of outward propagating shock wave trains. 
The hydrodynamic motions in these shocked layers are small (with Machnumbers close to unity), 
but their high temperatures account for the observed extra thermal line broadening that 
mimics a strong micro-`turbulence'. 

For $\alpha$~Sco A \citet{dej91} computed that these shock waves have average wavelengths 
of about twice the local density scale height, which is much longer than the correlation length for 
the small-scale density/temperature fluctuations affecting the population densities in the radiative 
transfer \citep[e.g.][]{heg00}. A microturbulent `filter function' provides 
the fraction of kinetic energy of the large-scale motion field at a given wavelength 
which contributes to the microturbulent broadening of a spectral line \citep{dej92}. 
Based on the line formation function, and the field of temperature fluctuations produced 
in the wakes of these shock trains, an integral function is obtained from which the 
atmospheric {\it quasi-microturbulence} can be computed. 
It is of note that in (limited) 3D simulations of compressible turbulence \citet{por97} 
demonstrated that shock waves develop from an initially uniform state which has been 
perturbed by randomly oriented, isentropic, sinusoidal sound waves. 
Many shock surfaces develop in the flow and intersect along lines 
from which slip surfaces, or sheets of vorticity emanate. 
Since a field of outward propagating sound waves is expected to be generated 
by the deeper convection zones in cool massive stars, these 3D shock wave 
collisions can develop into a large number of vortex tubes in higher layers.
Possibly they can produce the high microturbulence velocities we observe
in Betelgeuse's chromosphere?

Chromospheric Alv\'{e}n wave heating in $\alpha$~Ori was discussed by \citet{har84}.    
\citet{air00} recently discussed the results of 
2.5 MHD modeling work and concluded that a high mass-loss 
rate of $10^{-6}$~$M_{\odot}\,yr^{-1}$ can be explained 
if the magnetic field is several hundred gauss in the lower 
chromosphere. However, observational indications for 
such large magnetic field strengths have presently not 
been demonstrated.

Different types of detailed non-magnetic hydrodynamic calculations have 
been presented with respect to heating and dynamics of low-gravity atmospheres.
\citet{cun97} assumes the (stochastic) generation of acoustic waves by the 
subphotospheric convection zone that propagate into the chromosphere.
It is unclear how acoustic waves alone can create the global downfall
for the very extended chromosphere of Betelgeuse (more than seven times
$R_{\star}$), as we observe in our STIS spectra of 1998. 
\citet{bow88} computed the development of large-scale mass motions
in circumstellar environments which result from photospheric 
pulsations (or piston action) in AGB stars. These models do not propagate into 
a hotter chromosphere (as exists for Betelgeuse), and rely 
on radiation pressure on dust grains to accelerate the circumstellar 
gas. Betelgeuse's chromosphere is too hot for dust nucleation 
\citep{lob00}. We suggest that further hydrodynamic 
modeling should combine a cyclic acceleration of the deeper
atmosphere by photospheric pulsations which strongly enhance 
the atmospheric density scale height, with the generation of gravity 
waves that can produce sound waves and dissipating shocks at large distance 
from the surface. We think that the opportunity exists to 
compute such models for the very extended and dynamic chromosphere of 
Betelgeuse, but that has not yet been accomplished. 
   
Whether $\alpha$~Ori's extended chromosphere results 
from heating by viscous dissipation of turbulent energy, small-scale pressure waves 
and/or periodic photospheric accelerations, linked with the non-radial chromospheric 
movements we presently detect, can only be addressed when the source of these large-scale vibrations 
has been identified. 

\subsubsection{Chromospheric non-radial oscillation}

Semi-regular pulsation of stellar photospheres is traditionally attributed
to intermode beating of different radial pulsation frequencies, producing a long
`quasi'-periodicity, observed in Betelgeuse (SRc).
However, attempts by \citet{smi89} to fit 
its radial velocity (RV) curve with single- and double-mode sinusoids
did not provide any satisfactory explanation (for a discussion see
their Fig. 4). 

The long quasi-periodicity of light- and RV-curve of Betelgeuse 
is observed in other cool pulsating massive (Ia and Iab) supergiants as well. 
For example, for the yellow hypergiant $\rho$~Cas ($\rm Ia^{+}$), with 
$R_{\star}$$\simeq$400$\pm$100~$R_{\odot}$, \citet{lob94} demonstrated from an 
improved Baade-Wesselink pulsation test that photospheric {\em radial} pulsations 
must be discarded to combine $T_{\rm eff}$-changes with the co-eval 
RV- and light-curve, observed over a complete variability period.
Though the opacity, energetics, and atmospheric structure of $\rho$~Cas is 
likely quite different than $\alpha$~Ori 
due to differences in the effective temperatures and evolutionary state of these two stars.
\citet{gol84} noted the absence of a one-to-one correlation 
between short-term radial velocity and visual brightness
variations in Betelgeuse, which makes it 
apparent that these changes are not global in nature.   

Velocities associated with non-radial oscillation are expected to have 
a horizontal component as well. For adiabatic boundary conditions, the 
ratio of horizontal to vertical displacements is equal to 1 over the 
dimensionless pulsation frequency squared \citep{cox80}. For red supergiants 
this number is very small, suggesting that the component of vertical motions 
is very small. This is particularly true for gravity modes.
The small radial velocities we measure from the emission
line asymmetries across the chromospheric disk (not in excess of
2~$\rm km\,s^{-1}$ between simultaneous up- and downflows),
compared to the horizontal velocity required to extend the upflow in the 
lower chromosphere from the Western towards the Eastern hemisphere 
(thus at least over one stellar radius) between 1998 September and 
1999 March, appears to support the g-mode hypothesis. 

Hayes (1980, 1981, 1984) provided other observational indications for local 
non-radial pulsation of Betelgeuse's atmosphere.
Long-term monitoring of linear polarization changes reveals 
the presence of surface features which likely result from the 
waxing and waning of large-scale convective cells. Gravity modes 
with $l$=0 (radial motion) or $l$=1 (dipole motion) 
do not distort the star's spherical shape and can therefore be 
discounted to explain the changes of net polarization which occur 
over intervals of about one year. He also found that the changes 
of the $Q$ and $U$ Stokes parameters are not conform with polarization 
changes expected for a single {\em global} pulsation mode. 
However, {\em local} non-radial pulsations provide a means of 
producing the requisite asymmetry. The ordered changes of 
polarization position angle can be explained by the changes
in the orientation of a few large-scale surface elements 
(i.e. convection cells) which grow and fade over time, while 
migrating randomly across the stellar disk. Preliminary 
hydrodynamic 3D-simulations of giant granules in the 
photosphere of an entire red supergiant are presented in \citet{fre00}. 

Very high $l$-modes are not expected for supergiant pulsation because 
the wavelength of internal gravity waves propagating in these extended 
atmospheres should exceed the cut-off wavelength. The latter results from 
their radiative damping and the atmospheric curvature. 
Based on the diagnostic ($L$, $P$)-diagram \citet{dej91} 
pointed out that in the highest atmospheric levels of these stars only the 
very low-mode (long wavelength) gravity waves are possible. In deeper layers 
the damping cut-off wavelength becomes gradually shorter and higher gravity 
waves become possible. The more extended the atmosphere the lower the 
mode-number of gravity waves that may develop. 
In very cool stars, like Betelgeuse, for which the atmospheric pressure 
scale height is an appreciable fraction of the photospheric radius 
($\sim$10\% of $R_{\star}$), the convection region is well-developed 
and can extend up to optical depths as small as 0.3. However, it is 
known that gravity waves cannot propagate inside convectively stable 
regions where their minimum period (the Brunt-V\"{a}is\"{a}l\"{a} period) 
becomes real (and non-zero). Gravity waves may arise above the convective 
layers by penetrative (or overshoot) convective motions. They 
could be triggered in the very upper photospheric layers and propagate 
randomly into the lower chromosphere, causing the upflow motions we presently 
detect.

\citet{dje92} computed that a 440 day pulsation period for 
Betelgeuse can be interpreted as the period $P$ of photospheric internal gravity 
waves with a wavelength $L$ between $R_{\star}/3$ and $R_{\star}/30$. 
These low-order gravity waves propagate nearly horizontally through the photosphere. 
Non-radial pulsations of the deeper photosphere are a possible source
for g-modes that cause similar oscillations of the chromosphere, because the amplitude of 
the radial velocity curve measured in the optical at Oak Ridge Observatory 
\citep{dup99} indicates a redshift by 4 to 5 $\rm km\,s^{-1}$ during 
the monitoring period with STIS. An average collapse of the photosphere in the line of sight 
during this period is consistent with the mean downflow we observe for the larger chromospheric envelope. 
Localized subsonic upflow in the deeper chromospheric layers can result
from photospheric gravity waves that drive the complex chromospheric kinematics,
because the total velocity amplitude of $\sim$2~$\rm km\,s^{-1}$ we measure at the inner disk in 1998 September 
remains within the velocity range of $v=L/P=0.43$ to 4.3~$\rm km\,s^{-1}$, predicted for these waves. 
But note that this phase-velocity equals the velocity amplitude only for strictly horizontally propagating 
planar waves. In reality the wavevector of a gravity wave has a radial component as well, 
and the velocity amplitude will differ from the (group-)velocity of energy propagation.     

\citet{dje92} also noted that for Betelgeuse a period of 
$\sim$11 years, reported by \citet{dup90}, can result from gravity waves with a wavelength 
equal to the stellar circumference, or -what is the same- to photospheric non-radial 
pulsations with $l$=1. It is interesting to note that recent time-dependent 3D 
hydrodynamic simulations by \citet{jac99} 
of a complete red giant star show highly non-symmetric global dipolar 
flows within the convective layer, as well as distinct radial pulsations 
driven by non-linear interactions of acoustic waves with convective 
motions\footnote{for images see http://www.lcse.umn.edu/research/RedGiant/}. 
Spherical harmonics of the radial velocity are dominated by 
the dipolar ($l$=1) mode followed, at higher $l$-values, by a spectrum 
of modes similar to that of a Kolmogorov turbulence spectrum. They  
note that the simulated oscillations can be induced by the large-scale 
convective motions.  

A more definitive answer requires detailed hydrodynamic modeling and a stability analysis of our 
model chromosphere against time-dependent mechanical perturbations. The placement of further 
constraints on the temporal properties of Betelgeuse's chromospheric dynamics, 
based on semi-empirical modeling, would greatly benefit from observing 
near-UV chromospheric emission lines and unblended near-IR absorption lines that form deep in the 
photosphere, to track the velocity amplitudes and the detailed phase correlation of both atmospheric 
regions with sufficiently high temporal and spectral resolution, over at least a complete pulsation cycle. 
Future spectral studies should also comprise spatially resolved high-dispersion observations of the 
Na~{\sc i} or K~{\sc i} resonance lines to infer the velocity structure of the gas component in the 
(asymmetric) circumstellar dust shell, to test for possible correlations with the chromospheric kinematics
of this supergiant star.      

\section{Conclusions}
$i$. Spatially resolved high-resolution spectra of selected emission lines in 
Betelgeuse's near-UV spectrum reveal self-absorbed optically thick emission lines   
that display reversals in their component maxima when scanning across the chromospheric disk.
We identify four unsaturated emission lines, Si~{\sc i} $\lambda$2516.1, Fe~{\sc ii} $\lambda$2402.6,
Fe~{\sc ii} $\lambda$2868.9, and Al~{\sc ii}] $\lambda$2669.2, that display a very similar  
evolution of their component maxima and of line morphology during 
this monitoring period of 15 months.      
A prominent reversal is observed for these lines in 1998 September at two subsequent 
scan positions near the disk center. 

$ii$. We model with radiative transport calculations in NLTE and spherical geometry 
the detailed shape of the Si~{\sc i} $\lambda$2516 resonance line for different scan positions. 
The spatially resolved modeling of this component reversal reveals a mean velocity 
for the chromospheric fluid with opposite flow directions and a difference of velocity 
amplitude of $\sim$2~$\rm km\,s^{-1}$. This detailed modeling precludes a strictly 
radial mean velocity structure for the chromospheric kinematics.   

$iii$. The temporal behavior observed in many other emission lines 
during this monitoring period indicates a chromospheric oscillation phase
with global downflow between 1998 January and 1998 September. For the latter
observation date local upflow commenced in deeper chromospheric layers of the 
Western hemisphere, which extended further towards the Eastern hemisphere 
in March 1999. This global downflow is consistent with a redshift 
by 4 to 5 $\rm km\,s^{-1}$ in the disk-integrated photospheric 
radial velocity curve observed in the optical. 

$iv$. The detailed modeling of these spatially resolved Si~{\sc i} profiles requires a constant 
value for macro-broadening of 9$\pm$1~$\rm km\,s^{-1}$ at all scan positions. 
This implies a strongly isotropic and uniform large-scale velocity field across the stellar 
chromosphere, but which remains spatially unresolved with the 25~mas by 100~mas field of view.             
The modeling of the line profiles requires
highly supersonic microturbulence velocities in the chromosphere.

$v$. In general, we observe and measure with elaborate radiative transport 
calculations an increase of hydrodynamic velocities 
with a decrease of their length scales in the chromosphere of Betelgeuse. 
The very large-scale global chromospheric oscillation is non-radial 
and subsonic. The large-scale macroturbulence is isotropic with a velocity 
around the isothermal sound speed in the chromosphere. The small-scale 
microturbulent velocities are subsonic in the stellar photosphere 
and become highly supersonic in the chromosphere.    

\acknowledgments

We thank R. Kurucz for help with the LTE spectral synthesis 
calculations. R. Gilliland is gratefully acknowledged for assistance with the 
HST proposal of the STIS observations. We are grateful to the 
Armagh Observatory (UK) for providing additional computational support.
We thank the referee for useful comments to improve the clarity of the paper.
This research is supported in part by an STScI grant 
GO-5409.02-93A to the Smithsonian Astrophysical Observatory.

\clearpage

\begin{figure}
\figcaption[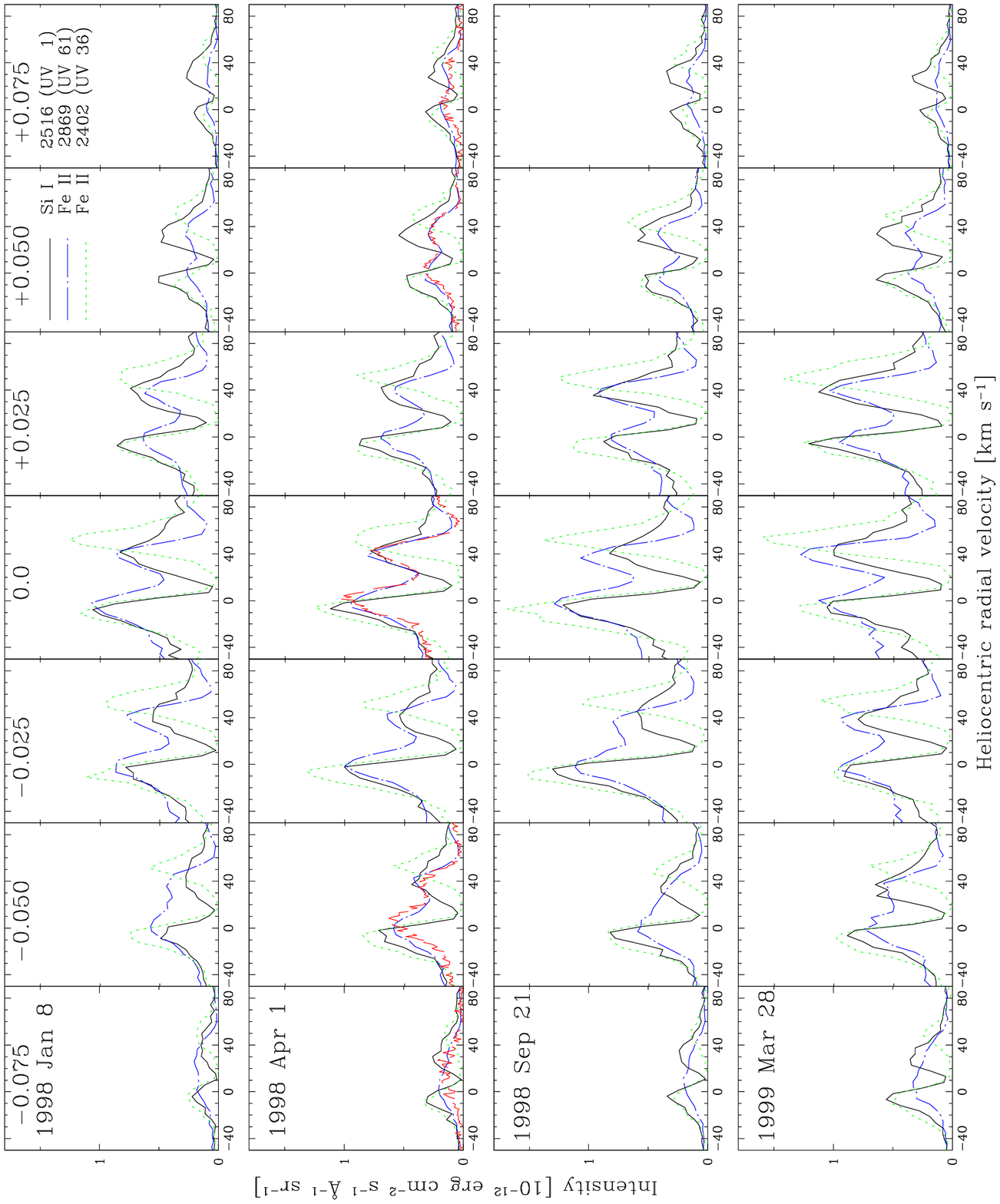]
{\normalsize Four spatially resolved raster scans of Betelgeuse's chromosphere, obtained with STIS
between 1998 January and 1999 March. The self-absorbed line profiles of Fe~{\sc ii} $\lambda$2868.9
({\it dash-dotted lines}), Si~{\sc i} $\lambda$2516.1 ({\it solid lines}), and Fe~{\sc ii} $\lambda$2402.6 ({\it dotted lines}), 
observed with medium-resolution and with an aperture size of 25~mas by 100~mas, are shown for 
three pointing off-sets of 25~mas left and right of intensity peakup (0.0). 
Note the remarkable reversals in the maxima of the emission line components 
from scan positions 0.0 to $+$0.025 in 1998 September. This reversal  
is modeled for the Si~{\sc i} line in this paper with detailed radiative transport calculations that reveal 
movements in opposite directions with subsonic mean velocities over the chromospheric 
line formation region. A high-dispersion observation of the Fe~{\sc ii} $\lambda$2868.9 line  
is shown by the dashed lines in the scan of 1998 April. These line profiles are obtained 
through an aperture of 63~mas by 200~mas for two off-sets of 63~mas, left and right of 
peakup. The absolute intensity of these profiles has been scaled down for comparison with the 
medium-resolution scan. The relative intensity of the line components and the line 
shape do not alter with higher spectral resolution, showing that the medium-resolution 
observations are useful for the detailed spatially resolved modeling in this paper.\label{fig1}}
\end{figure}
\begin{figure}
\plotone{fig1.ps}
\end{figure}

\clearpage

\begin{figure}
\figcaption[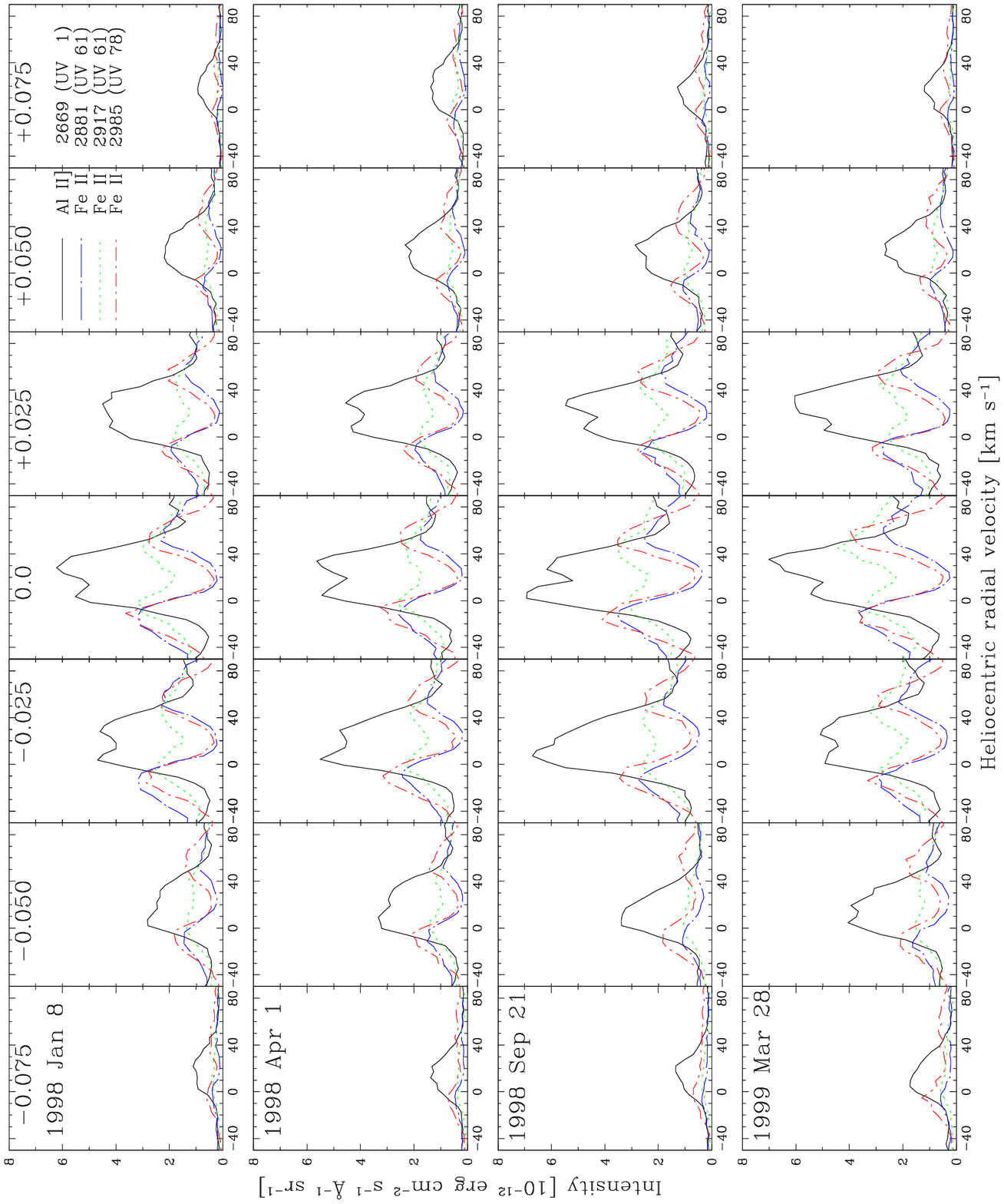]
{\normalsize Medium-resolution scans for Al~{\sc ii}] $\lambda$2669.1 ({\it solid lines}), 
Fe~{\sc ii} $\lambda$2917.4 ({\it dotted lines}), Fe~{\sc ii} $\lambda$2880.7 ({\it long dash-dotted lines}),
and Fe~{\sc ii} $\lambda$2984.8 ({\it short dash-dotted lines}). The Al~{\sc ii} line also displays the
reversal of its component maxima in 1998 September, similar as for the lines in Figure~1
(see text). Line profiles are shown un-smoothed.\label{fig2}}
\end{figure}
\begin{figure}
\plotone{fig2.ps}
\end{figure}

\clearpage

\begin{figure}
\figcaption[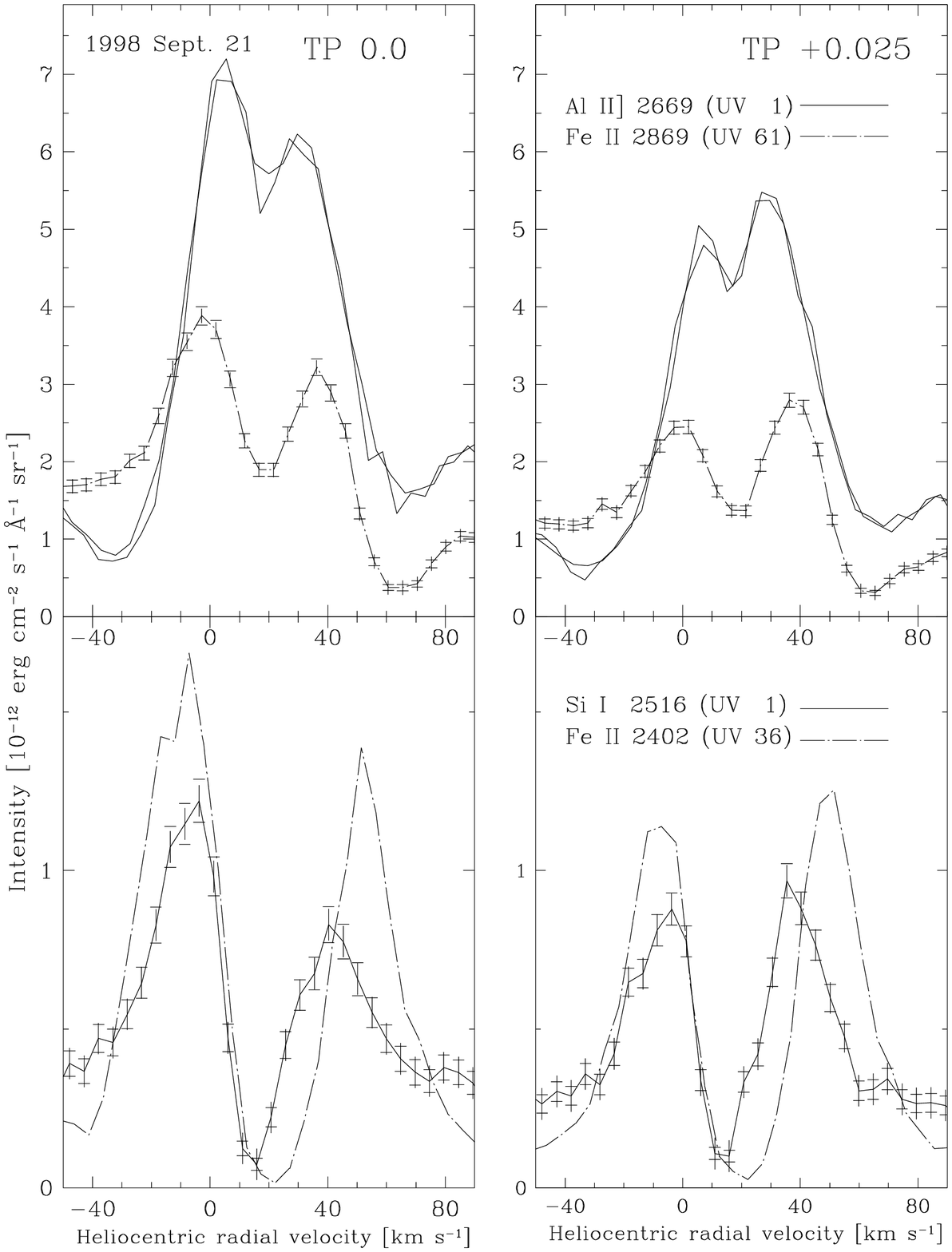]
{\normalsize Profile changes in unsaturated self-reversed chromospheric emission lines reveal
systematic reversals of the emission component maxima near the disk center (TP 0.0, {\it panels left}; 
TP~$+$0.025, {\it panels right}) for the spatial scan of 1998 September~21.
The Al~{\sc ii}] $\lambda$2669 line for these observations
is shown for two overlapping echelle orders ({\it solid lines in upper panels}), 
which reveals that the instrumental noise remains small. These spatial variations of the component intensities
exceed the errorbars provided by the STIS calibration pipeline, 
shown for the Fe~{\sc ii} line ({\it dash-dotted lines in upper panels}) and the Si~{\sc i} line
({\it solid lines in lower panels}).\label{fig3}}
\end{figure}

\clearpage

\begin{figure}
\plotone{fig3.ps}
\end{figure}

\clearpage

\begin{figure}
\figcaption[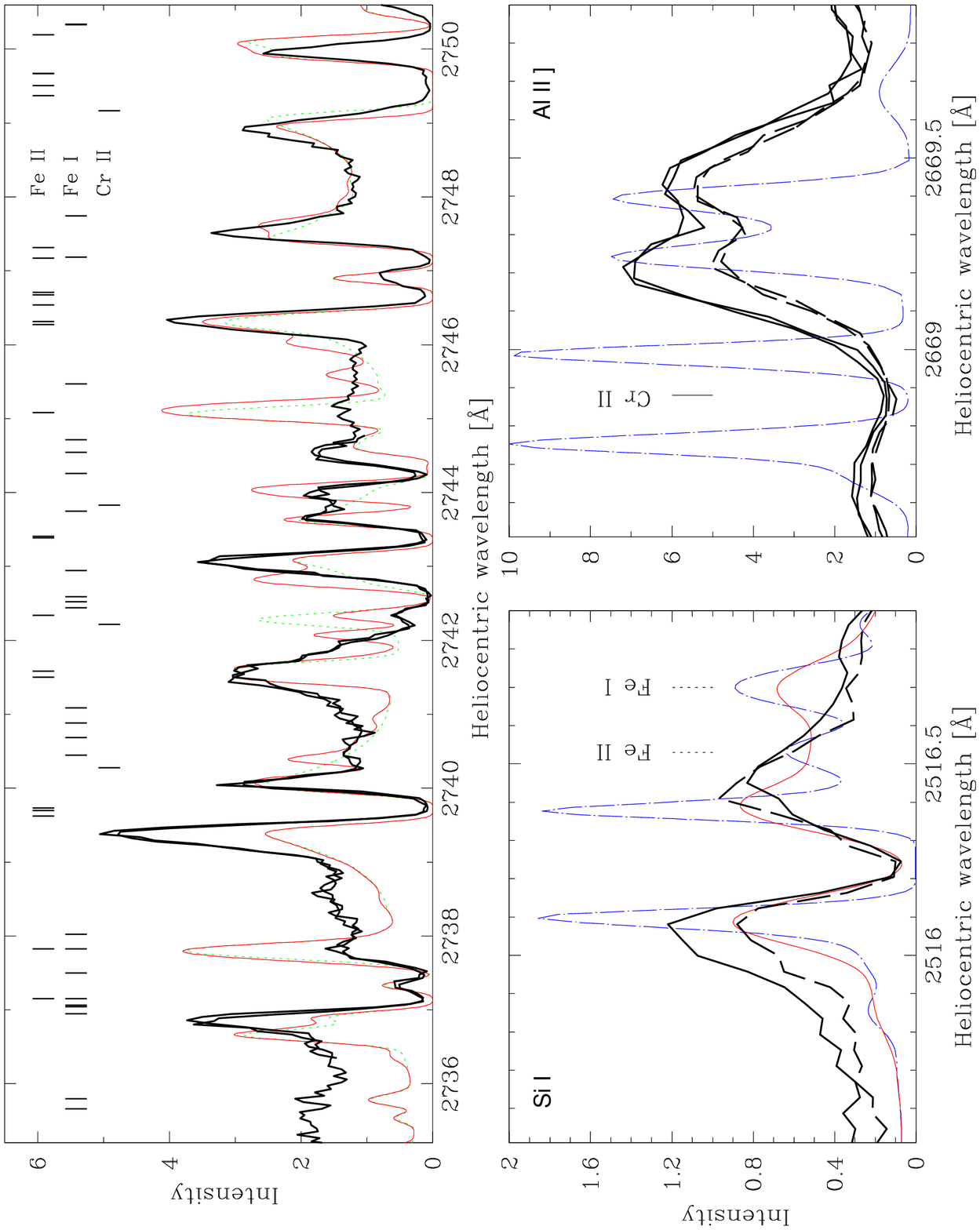]
{\normalsize {\it Upper panel:} Portion of Betelgeuse's near-UV spectrum observed with STIS 
for intensity peakup (TP 0.0) with medium-resolution on 1998 September 21 ({\it bold solid line}).
The LTE synthesis ({\it thin solid line}) with a hydrostatic model of the chromosphere (Lobel \& Dupree 2000) 
shows that the deep absorption cores result from strong scattering cores of Fe~{\sc ii} lines, 
marked with the vertical lines in the stellar rest frame (top). Note that only Fe~{\sc i}, Fe~{\sc ii}, and Cr~{\sc ii}
lines are shown, although the synthesis also includes other atomic species. A synthesis which includes 
only iron lines ({\it dotted line}) already matches this observation very closely. 
It shows that the UV emission spectrum of the star mainly results from the complex 
term structure of iron, which causes many blends of very optically thick line transitions in this extended 
chromosphere. {\it Lower panels:} An LTE synthesis (static atmosphere) of the Si~{\sc i} $\lambda$2516 line shape 
({\it thin dash-dotted line}) matches the observed profile of 1998 September ({\it bold solid line} for TP~0.0; 
{\it dashed solid line} for TP~$+$0.025), after convoluting with the instrumental resolution 
and macrobroadening of 9~$\rm km\,s^{-1}$. The two iron lines at the indicated wavelengths are
not observed. The profiles computed for the Si~{\sc i} and the Al~{\sc ii} line
are symmetric, which indicates that their observed spatial component reversals are not influenced 
by chromospheric blends.\label{fig4}}
\end{figure}
\begin{figure}
\plotone{fig4.ps}
\end{figure}

\clearpage

\begin{figure}
\figcaption[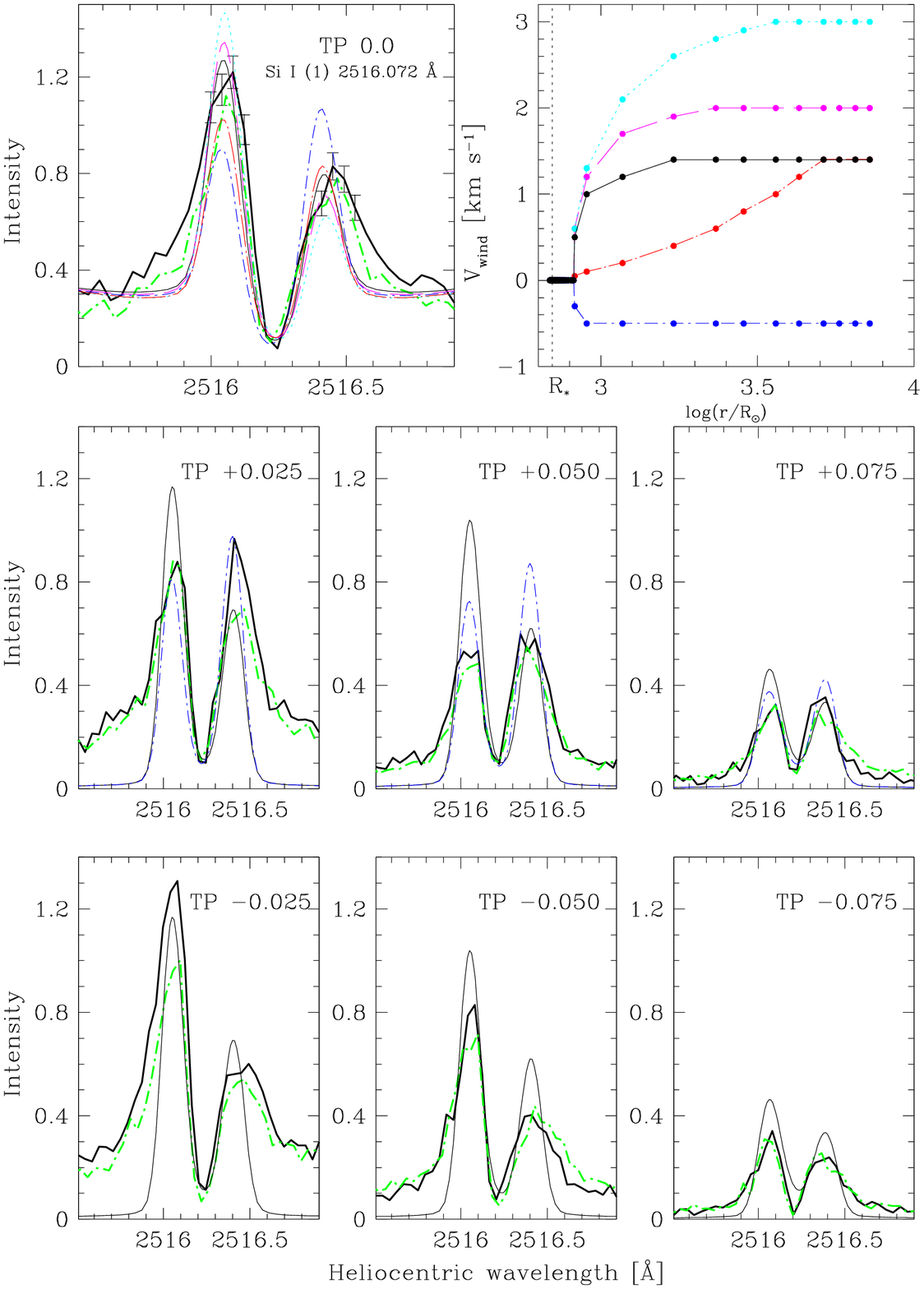]
{\normalsize Best NLTE fits ({\it thin solid lines}) to the Si~{\sc i} $\lambda$2516 line, observed in 
1998 September ({\it bold solid lines}) and in 1998 April ({\it bold dash-dotted lines}),
are computed in spherical geometry with a semi-empirical model of the mean 
velocity structure in the chromosphere, shown in the upper right-hand panel ({\it solid line}). 
This model with downflow reproduces the relative intensity of the emission line components 
at TP 0.0, but does not fit the intensity reversal observed at TP $+$0.025 in 1998 September. 
The component ratios for the positive scan positions are best fit ({\it thin dash-dotted lines}) 
with a model of subsonic outflow ({\it dash-dotted line in upper right-hand panel}). 
It reveals that the chromosphere assumes a non-radial velocity structure in 1998 September, 
for which the spatially resolved modeling measures a difference of flow velocity in opposite 
directions of $\sim$2~$\rm km\,s^{-1}$. Notice how these spatially resolved calculations 
also reproduce the reduction in the component ratios, caused by the projection effect, 
when scanning towards the limb (compare lower panels).\label{fig5}}
\end{figure}

\clearpage

\begin{figure}
\plotone{fig5.ps}
\end{figure}

\clearpage

\begin{figure}
\figcaption[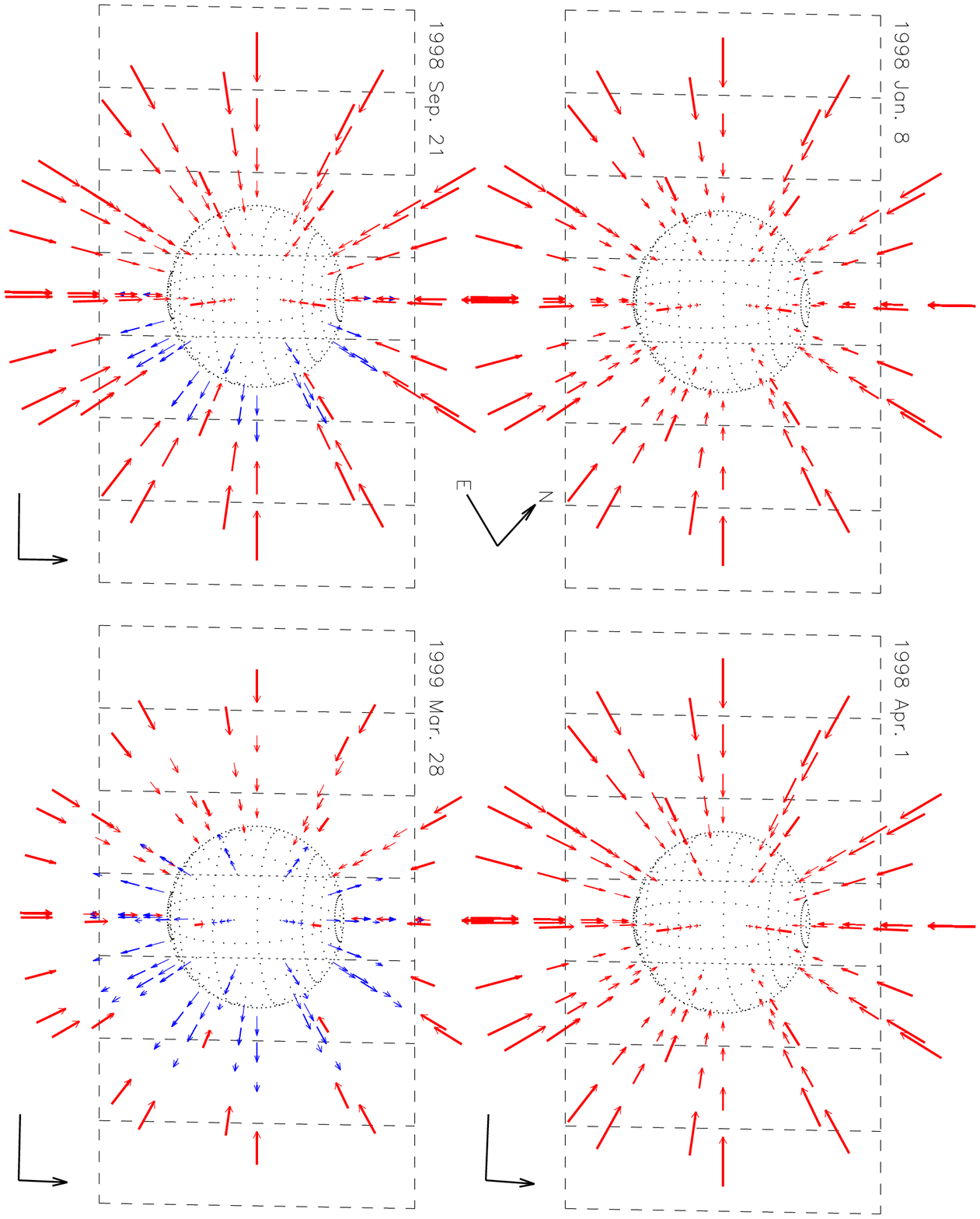]
{\normalsize Schematic 3D representation of the fluid movements in Betelgeuse's extended chromosphere, 
based on the spatially resolved semi-empirical modeling of the Si~{\sc i} $\lambda$2516 line profile
in the STIS raster scans, obtained between 1998 January and 1999 March. Other more optically thick lines define the 
dynamics at high levels. The global downflow observed for the larger chromospheric envelope in 1998 January and 1998 April, 
reverses into subsonic upflow in 1998 September for the lower chromospheric layers. 
It reproduces the reversal in the component maxima of this line, observed for different scan
positions of the aperture ({\it dashed lines}) near the inner chromospheric disk.
The photospheric radius is drawn by the dotted circles. 
The outflow enhances in 1999 March when more symmetric profiles are observed, extending 
further towards the Eastern (left-hand) front hemisphere. The detailed modeling reveals a local non-radial 
oscillation of the chromosphere.\label{fig6}}
\end{figure}
\begin{figure}
\plotone{fig6.ps}
\end{figure}

\end{document}